\documentclass[12pt,preprint]{aastex}

\slugcomment{draft version \today, }

\shorttitle{Direct Time Variable Radio Images For M87}
\shortauthors{Nagakura et al.}

\begin{document}

\title{
 Direct time radio variability induced by \\
 non-axisymmetric standing accretion shock instability: \\
 implications for M87}

\author{Hiroki Nagakura\altaffilmark{1}, Rohta Takahashi\altaffilmark{2}} 

\altaffiltext{1}{Department of Science and Engineering,
 Waseda University, 3-4-1 Okubo, Shinjuku, Tokyo 169-8555, Japan}
\altaffiltext{2}{Cosmic Radiation Laboratory,
 the Institute of Physical and Chemical Research (RIKEN) 2-1 Hirosawa, Wako, Saitama 351-0198, Japan}

\begin{abstract}
 We show for the first time the direct time variable radio images
 in the context of shocked accretion flows
 around a black hole under the general relativistic treatment
 of both hydrodynamics and radiation transfer.
 Time variability around a black hole can be induced by the non-axisymmetric
 standing accretion shock instability (namely Black Hole SASI).
 Since the spiral arm shock waves generate
 the density and temperature waves at the post shock region,
 they cause time variability in the black hole vicinity.
 Based on our dynamical simulations,
 we discuss a possibility of detection for the time variable radio images
 of M87 by the future space telescope
 {\it VSOP2/ASTRO-G} satellite.
 The most luminous part of the images is predicted to be near 15 Schwarzschild
 radii for some snapshots.
 We show that our results are consistent with existing observational data
 such as time-averaged radio spectra,
 VLBA images and variability timescale for M87.
 We also discuss observations 
 of M87 with millimeter and sub-millimeter interferometers. 
\end{abstract}



\section{Introduction}
Elucidating the nature of the strong field of gravity around black holes and the physics of the accreting plasma plunging into black holes is a major challenge in physics and astrophysics. Thanks to rapid progress in interferometric techniques from the radio to sub-millimeter wavelengths, within the next decade the regions in the vicinity of the massive black holes in e.g. our Galactic center, Sgr A*, and the elliptical galaxy, M87, whose apparent sizes are few tens of micro-arc seconds, will be directly imaged \citep{fma00, bl06, h07, m07, d08, bl09}. The {\it VSOP-2/ASTRO-G} project, which is a space-VLBI (Very Large Baseline Interferometer) mission, will achieve a spatial resolution of 38 $\mu$arc seconds at 43 GHz, corresponding to $\sim 20 GM/c^2$ for the black hole in M87 for a black hole mass $M=3.2 \times 10^9  M_\odot$ and distance $D=16 $Mpc \citep{macchetto1997,m99,tsuboi2008} \footnote{Since the central region around the black hole in Sgr A* is smeared out by intersteller scattering at 43 GHz \citep{shen2005}, the central region will not be spatially resolved.}. Here, $G$ is the gravitational radius and $c$ is the speed of light.  

In the central region of M87, the strong jets are observed from radio to X-ray/gamma-ray energies, and high resolution images obtained by the VLBA (Very Long Baseline Array) show the bright region from which jets are launched \citep{junor1999,dodson2006,walker08}. The observational data in M87 generally show time-variability features with timescales ranging from days for TeV gamma-rays \citep{Aharonian2006,Albert2008} to months for radio, optical and X-ray bands \citep{harris1997,perlman2003}. These timescales correspond to, or are larger than, the Keplarian time-scale at the innermost stable circular orbit (ISCO) of the black hole in M87, from a few days to one week depending on the black hole spin. Theoretically, the observational variability has been explained based on the accretion disk - jets systems \citep{chakra1993,dimatteo2003,becker08,bl09,mandal2008}. In particular, \citet{chakra1993} have pointed out that variations in emission from accretion disks could be explained by non-axisymmetric spiral shock waves. \citet{mandal2008} also mentioned that the standing shock model can explain the timescale of a few months by comparisons with the free fall timescale in the sub-Keplerian flows based on the steady state analysis. As in the examples given above, the standing shock waves are important for the time variable emissions in the vicinity of black holes. They have attracted much attention from many researchers in this field \citep{nak94,nak95,das03a,das03b,chakra2004,oku07}.

 On the other hand, recent theoretical studies show that the accretion shock waves are generally unstable against asymmetric perturbations (namely standing accretion shock instability or SASI), and they might play an important role for the quasi-periodic oscillations in stellar mass black hole candidates \citep{nagakura2008}, or for supernova explosions \citep{blondi03,scheck08,iwakami08,marek09}. According to \citet{molteni1999} and also \citet{nagakura2008,nagakura2009}, standing shock waves are deformed by non-axisymmetric perturbations, and eventually form spiral arm structures. Although the linear analysis shows that the shock waves are unstable against non-axisymmetric perturbations \citep{nagakura2008,nagakura2009}, these shock waves keep rotating and oscillating in the non-linear phase, while the spiral density and temperature fluctuations from shocks are successively plunging into a black hole. These shock waves are very different from axisymmetric ones. According to linear analysis and dynamical simulations by \citet{nagakura2008,nagakura2009}, the rotation and oscillation periods depend on the location and strength of the shock waves. They may be related with the instability mechanism which still remains elusive. The strongest candidate for the mechanism of SASI in black hole accretion (namely Black Hole SASI) is the Papaloizou-Pringle type instability, which is induced by acoustic-acoustic cycles between a shock radius and an inner sonic surface (or possibly a corotation radius) \citep{Gu03,Gu06,nagakura2008,nagakura2009}.

 The dynamics of Black Hole SASI are affected by the strong gravitational potential. This suggests that if the observational evidence of Black Hole SASI can be obtained, it can be used as a diagnostic for the dynamics of plasma around black holes. Thus, it is necessary to investigate the possibility of detection. In this paper, we perform hydrodynamic simulations of the Black Hole SASI and the radiative transfer by taking into account the general relativistic effects around a Schwarzschild black hole in order to investigate the observational feasibility of the time variable features from M87. We show that our results are consistent with the observational features of M87 such as energy spectrum, image morphology taken by the VLBA, and variability with timescales of a few months. We also show that the time variability induced by spiral waves swallowed by the black hole will be observed by the high detection sensitivity and time resolution of the {\it VSOP2/ASTRO-G} satellite.

 This paper is organized as follows. In section 2, we explain the methods and models in this work. The numerical results and comparisons with M87 are given in section 3. Finally we summarize and conclude this paper in section 4. 

\section{Methods and Models}
 In this paper, we consider non-axisymmetric shocked accretion flows
 onto a Schwarzschild black hole. 
 We briefly explain the numerical procedure and models in this section. 
  
\subsection{Hydrodynamics}
 The detailed procedure for the hydrodynamic simulations is
 described in \citet{nagakura2008}.
 We employ a $\Gamma$-law equation of state (EOS) $p = \left(\Gamma-1\right)\rho_0 \epsilon$,
 where $\Gamma$, $p$ and $\epsilon$ are the adiabatic index, pressure,
 and specific internal energy, respectively.
 In this paper, we use the time-dependent
 simulation data for the standard model (M1 model) in \citep{nagakura2008}.
 The initial shocked accretion flows are set as steady solutions
 which are obtained by combinations
 of the adiabatic index and injection parameters
 such as Bernoulli constant ($E$) and specific angular momentum ($\lambda$).
 We set these parameters as
 $\Gamma = 4/3$, $E/c^2=1.004$, and $\lambda \times c / G M = 3.43$, respectively.
 As shown in \citet{lu1985,nagakura2008},
 these conditions allow multiple sonic points as well as
 two-possible shocked accretion flow.
 Since it is well known that the inner shock wave is already
 unstable against axisymmetric perturbations in the black hole accretion flow,
 the outer shock solution comes into existence in a realistic system.
 Thus, we only consider the outer shock
 which is located at $r_{\rm sh}=16.1GM/{c^2}$ in this model.

 According to linear analysis \citep{nagakura2008,nagakura2009},
 the outer shock wave is unstable against non-axisymmetric perturbations
 in the linear regime.
 We have computed this system in the equatorial plane with a multi-dimensional
 general relativistic hydrodynamical code which is based on
 the high-resolution central scheme, originally proposed by \citet{kur2000}.
 Actually, our simulations have shown that
 after non-axisymmetric perturbations are added,
 the initial axisymmetric shock wave starts to deform
 and quasi-steady spiral arm structure was realized in fully non-linear regime.
 The dominant mode for deformed shock waves in non-linear regime
 is a lower mode ($m=1$~or~$2$ mode)
 which keeps rotating and oscillating axisymmetrically and semipermanently
 until the outer boundary conditions are changed.
 We refer the reader for more details of Black Hole SASI hydrodynamics
 to \citep{nagakura2008,nagakura2009}.
 In this paper, since we are interested in only the non-linear regime,
 we use these data from enough runs on these simulations
 ($t \sim 3 \times 10^3 G M /c^3$)
 to the end of our simulations ($t \sim 3 \times 10^4 G M /c^3$)
 for further radiation transfer calculations.

\subsection{Radiation Transfer}
By using numerical simulation data described above, we perform general relativistic radiative transfer calculations by using the ray-tracing method including the usual special and the general relativistic effects such as Doppler boosting, gravitational redshifts and light bending. We assume the mass $M$, the distance $D$ of the black hole in M87 as $M=3.2\times 10^9M_\odot$ \citep{m97}, $D=16$ Mpc \citep{m99}, respectively. Although in this study, we assume the Schwarzschild metric, we use the code of null geodesics in Kerr spacetime \citep{t04,tw07}. For $r$ and $\theta$ components of the null geodesics, we use semi-analytic expressions given by \cite{rb91}, and $t$ and $\phi$ components are numerically solved by the Romberg integration. In this paper, we only consider thermal synchrotron emissivity and absorption in the accretion flow. This is because previous studies have shown that the observed spectrum at 43 GHz can be fitted by a thermal synchrotron model \citep{dimatteo2003,wang08, li2009,bl09}. In this study, we assume the emissivity $j_\nu$ and the absorption coefficient $\alpha_\nu$ of the thermal synchrotron radiation of the RIAF model \citep{m00, yqn03} and calculate the intensity $I_\nu$ observed at 43 GHz seen by a distant observer, by integration along the null geodesics as $I_\nu=\int e^{-\tau_\nu}g^3 j_\nu dl$ where $dl$ is the proper length differential along the geodesics, $\tau_\nu\equiv \int \alpha_\nu dl$ is the optical depth and $g\equiv \nu_{\rm obs}/\nu$ \citep{fw04,bl09}.  Here, $\nu_{\rm obs}$ is the observed frequency and $\nu$ is the frequency at the local rest frame.  When calculating this $g$-factor, we use the simulation data of the four-velocity and the special and general relativistic effects are included. The emissivity and the absorption coefficients are calculated from the density and the temperature of hydrodynamic simulations. We assume a vertical distribution of the number density and the temperature of the accretion flow obeying the exponential or Gaussian distributions with a Full Width at Half Maximum (FWHM) of $r \sin (\pi/4)$ for simplicity. Since the density profile of the hydrodynamic simulations are normalized by mass accretion rate, we choose the mass accretion rate so as to reproduce the observed spectrum. Our numerical simulations do not solve the electron temperature and the magnetic field which are required to calculate the observed images based on the synchrotron radiation. In this study, the electron temperature and magnetic fields are treated as in \cite{bl09} and we assume plasma beta $\beta = 10$ in all regions. 

\begin{figure}
\hspace{-10mm}
\epsscale{1.0}
\plotone{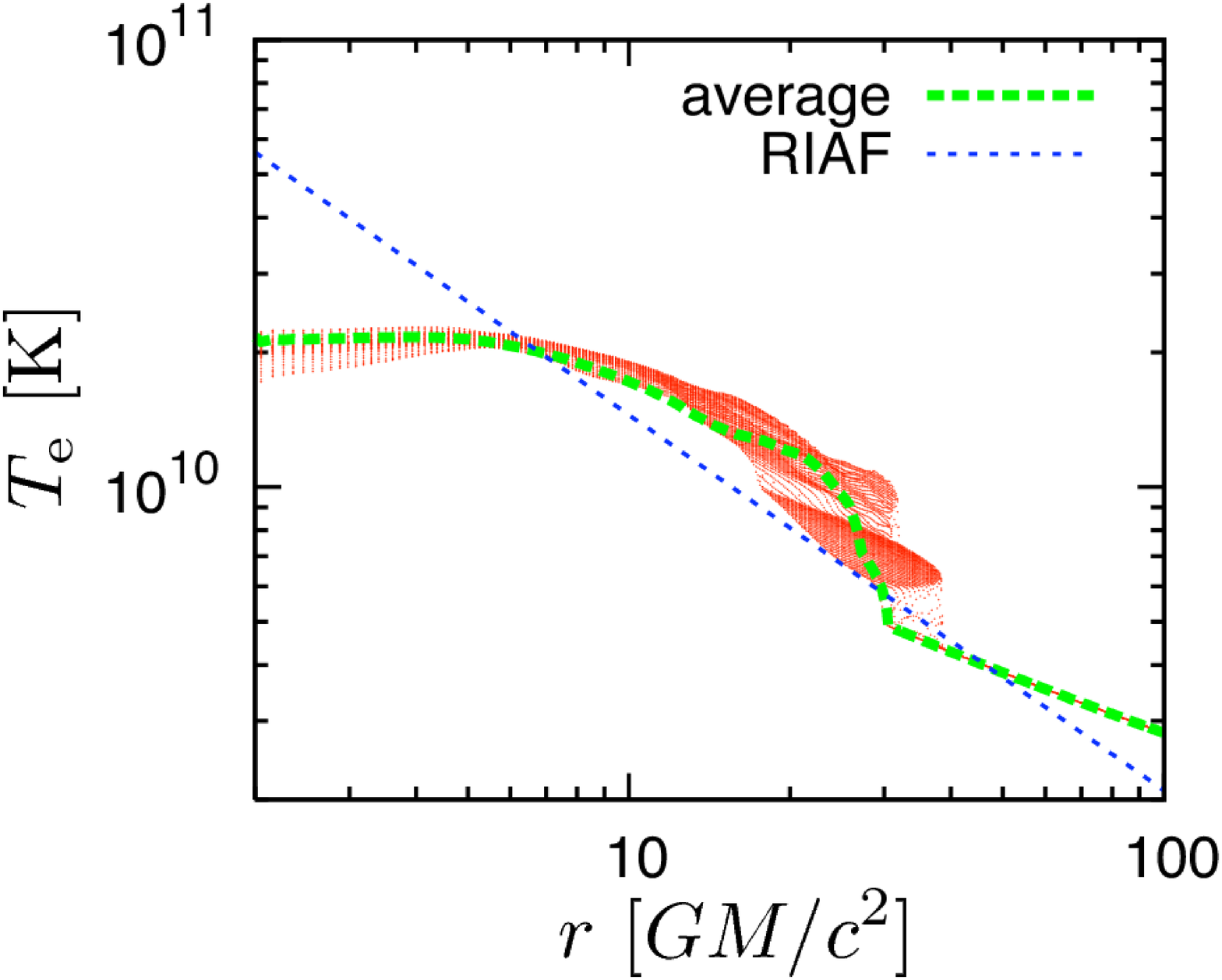}
\caption{
The radial profiles of the electron temperature $T_e$ [K] for one snapshot ({\it dots}) and the time-averaged data ({\it thick dashed line}) for our simulation data. We also show the electron temperature profile for the RIAF model used in \cite{bl09} ({\it dotted line}). 
\label{fig:te}}
\end{figure}

 By using the simulation data and assumptions denoted above, we fit the observed radio spectrum of M87. In Figure~\ref{fig:te}, we show the radial profiles of the electron temperature $T_e$ [K] for one snapshot ({\it dots}) and the time-averaged data ({\it thick dashed line}) for our simulation data. Note that since electron temperature depends on the azimuthal direction for one snapshot, there are several dots at the same radius in this figure. We also show in Figure~\ref{fig:te} the electron temperature profile for the RIAF model used in \cite{bl09} ({\it dotted line}). Although the electron temperature profiles used in this paper become nearly constant near the black hole compared to the RIAF model, the absolute values of the electron temperatures are roughly similar to those of the RIAF model. In Figure \ref{fig:sed}, we show the results of the spectral fitting of our models ({\it solid lines}) to the observed radio spectra of M87 ({\it cross signs}).  Here, we use the simulation data of eight snapshots and the observational data in \cite{bsh91} ($1.49\times10^{10}$ Hz), \cite{sj86} ($2.18\times10^{10}$ Hz), \cite{dfd96} ($8.90\times10^{10}$ Hz) and \cite{b92} ($10^{11}$ Hz). At 43 GHz, the observed spectrum can be fitted by assuming synchrotron emissivity due to thermal electrons \citep{dimatteo2003,wang08,bl09,li2009}. After calculating the images of the intensity $I_\nu$ observed by the distant observer, we take into account the effects of the spatial resolution at 43 GHz of the {\it VSOP-2/ASTRO-G} satellite by using the $u$-$v$ coverage data created by the simulation tool, ARIS \citep{a07} and calculate the map of the spatially-smoothed intensity $I_\nu^{\rm VSOP-2}$ as $I_\nu^{\rm VSOP-2}(x,y)=\int\int e^{2\pi i(xu+yv)} S(u,v) V(u,v) du dv$ where $V(u,v)$ is the visibility obtained by Fourier transformation of $I_\nu$, and $S(u,v)$ is the sampling function calculated from the $u$-$v$ coverage of the {\it VSOP-2/ASTRO-G} observations.  By performing the calculations for the simulation data, we finally create the time-variable images of the central region of M87 observed at 43 GHz by the {\it VSOP-2/ASTRO-G} satellite.

\subsection{Summarizing Our Assumptions}
 Since we impose many simplifications in this study,
 here we summarize the important assumptions in order for the reader
 to understand our models.

 Firstly, we neglect the magnetic field in the gas dynamics,
 although the radiation is assumed to be due to synchrotron emission.
 This assumption is inconsistent, but we neglect here for the following reasons.
 In this study, we consider the 'observational feasibility' of Black Hole SASI
 which does not depend strongly on the effects of magnetic field
 to gas dynamics.
 The details of gas dynamics should be analyzed
 under the general relativistic magneto-hydrodynamic (namely GRMHD) treatment
 in black hole space times \citep{mackinney04}.
 On the other hand, the effects of magnetic field to Black Hole SASI dynamics
 are unclear at the present time. Addressing these issues are beyond the
 scope of this paper.
 We are interested in how the magneto-hydrodynamic turbulence induced by MRI \citep{balbu1991}
 affects SASI, but these are deferred to future work.

\begin{figure}
\hspace{-10mm}
\epsscale{1.0}
\plotone{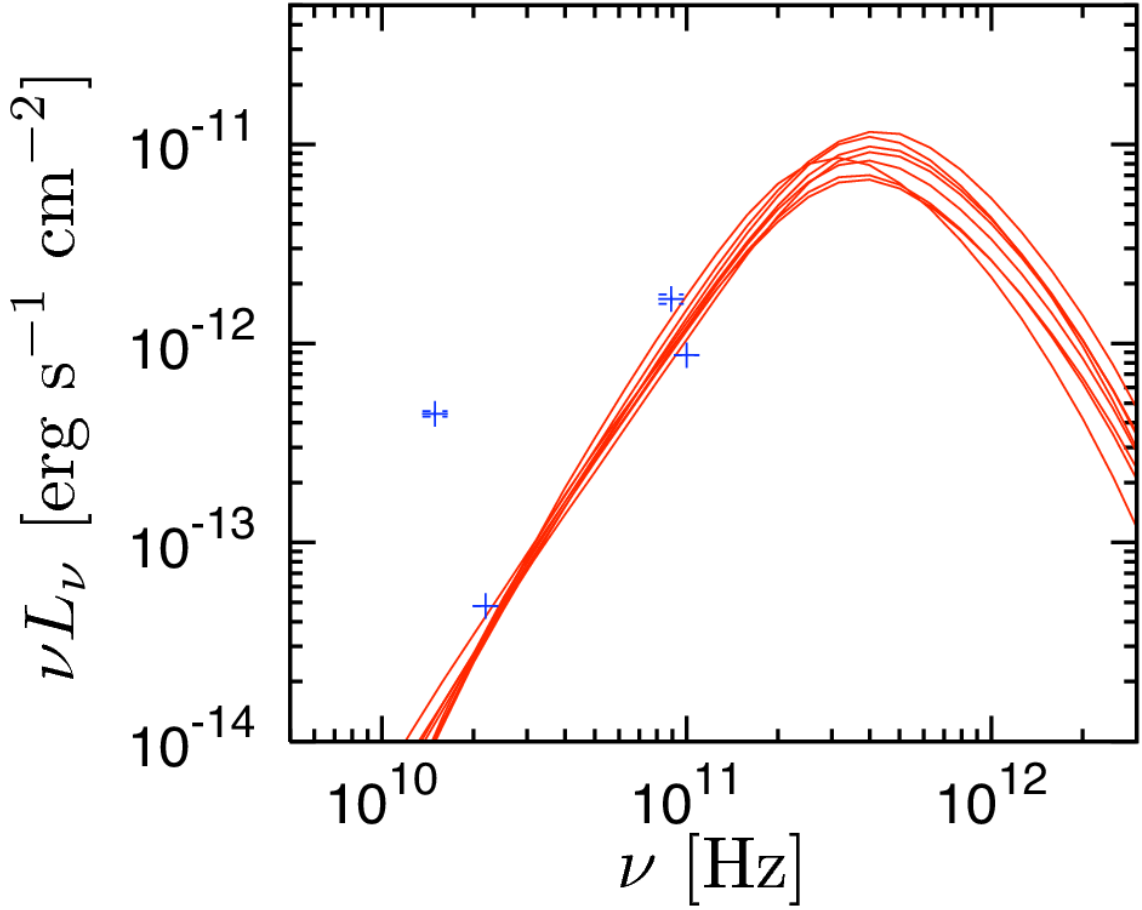}
\caption{
The results of the spectral fitting of our models ({\it solid lines}) to the observed 
radio fluxes of M87 ({\it cross signs}).  Here, we use the simulation data of 
eight snapshots and the observational data in \cite{bsh91} ($1.49\times10^{10}$ Hz), 
\cite{sj86} ($2.18\times10^{10}$ Hz), \cite{dfd96} ($8.90\times10^{10}$ Hz) and \cite{b92} ($10^{11}$ Hz). 
See also \citet{Ho1999} for additional data near 43 GHz. 
\label{fig:sed}}
\end{figure}

Secondly, we consider the thermal synchrotron emissivity and the absorption in the accretion flow. In the RIAF model \citep{yqn03}, synchrotron radiation due to thermal and non-thermal electrons is usually assumed. In this study, as shown in Figure \ref{fig:sed}, the observed energy spectrum can be fit by the thermal synchrotron at 43 GHz, thus we neglect non-thermal components in this study. At 43 GHz, the outflow components can also be negligible compared to the disk component \citep{junor1999,dimatteo2003,dodson2006}, although it should be included in the other energy bands \citep{bl09,li2009}. Actually, in our calculations, we can fit the observational data near 43 GHz, only using the thermal synchrotron components from accretion flows (see, Figure \ref{fig:sed}). 

Thirdly, our hydrodynamic simulations do not include meridional plane. For this reason, we assume the vertical distributions of the matter as the Gaussian or exponential profile for now. The consistent treatment of vertical structure for accretion flows is a major challenge for many years, since the multidimensional singular surface appears in  the steady accretion flows onto black hole (but see e.g. \citet{Sincell1997,Salmeron2007}). The steady solution is necessary as the initial conditions for the investigation SASI in vertical dynamics. We will address these issues in order to investigate more detailed time variability from Black Hole SASI.

\section{Results and Implications for M87}

\begin{figure*}
\epsscale{1.0}
\plotone{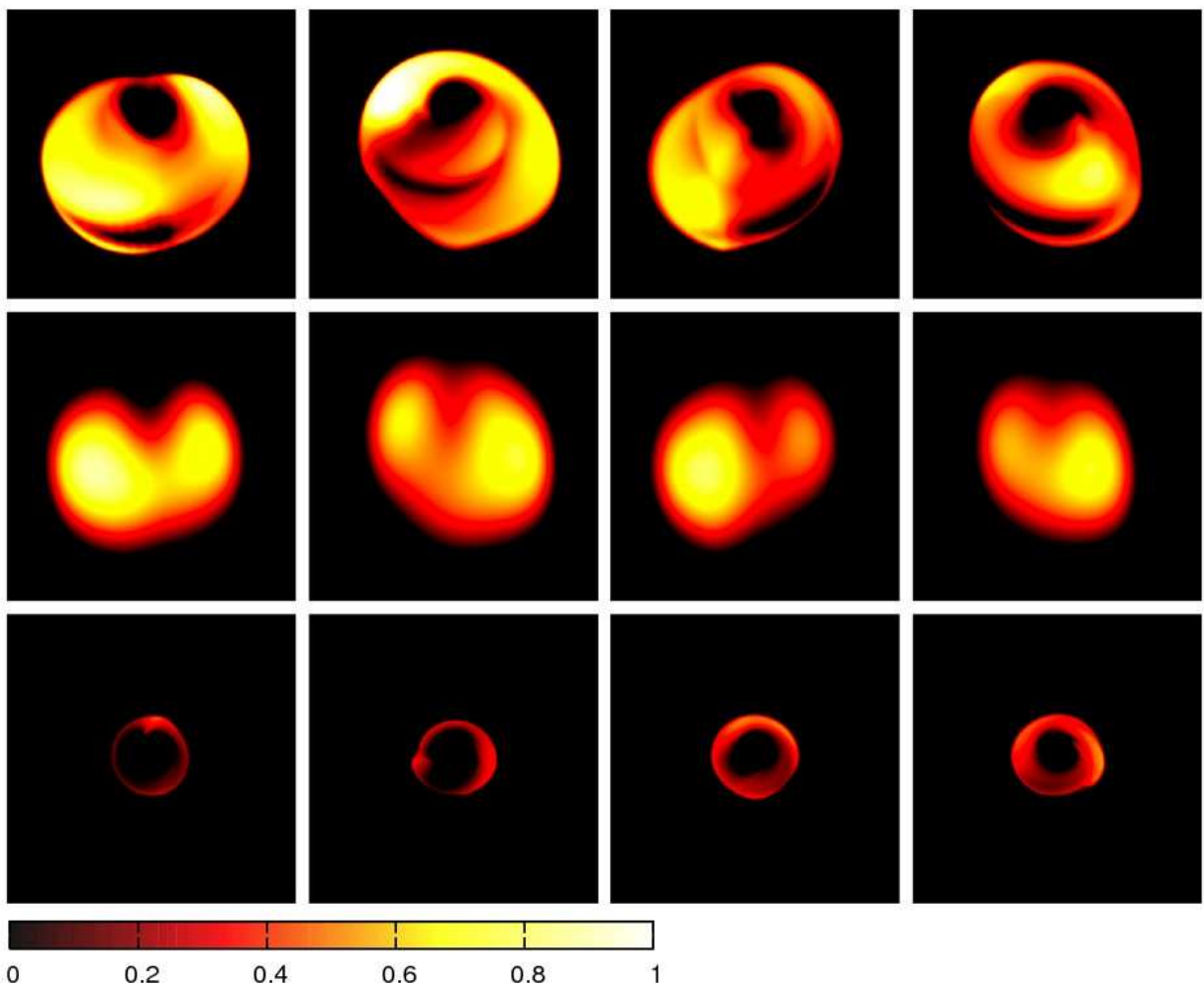}
\caption{
The simulated intensity distributions of the spiral waves caused by the SASI around a black hole ({\it top row}) and the images smeared by the spatial resolution of the {\it VSOP-2/ASTRO-G} observations at 43 GHz ({\it middle row}).  From left to right panels, the time evolution of the images are shown of time intervals of  $2.8\times10^6$ [s] $\sim$ one month for the black hole in M87.  The size of the images is $(80 GM/c^2)^2$. The bottom panels show the images at 345 GHz for the same model as the top panels. The viewing angle is assumed to be $i=30^\circ$. For hydrostatic equilibrium, an exponential profile of the vertical density structure is assumed. 
\label{fig:image}}
\end{figure*}

\begin{figure*}
\epsscale{1.0}
\plotone{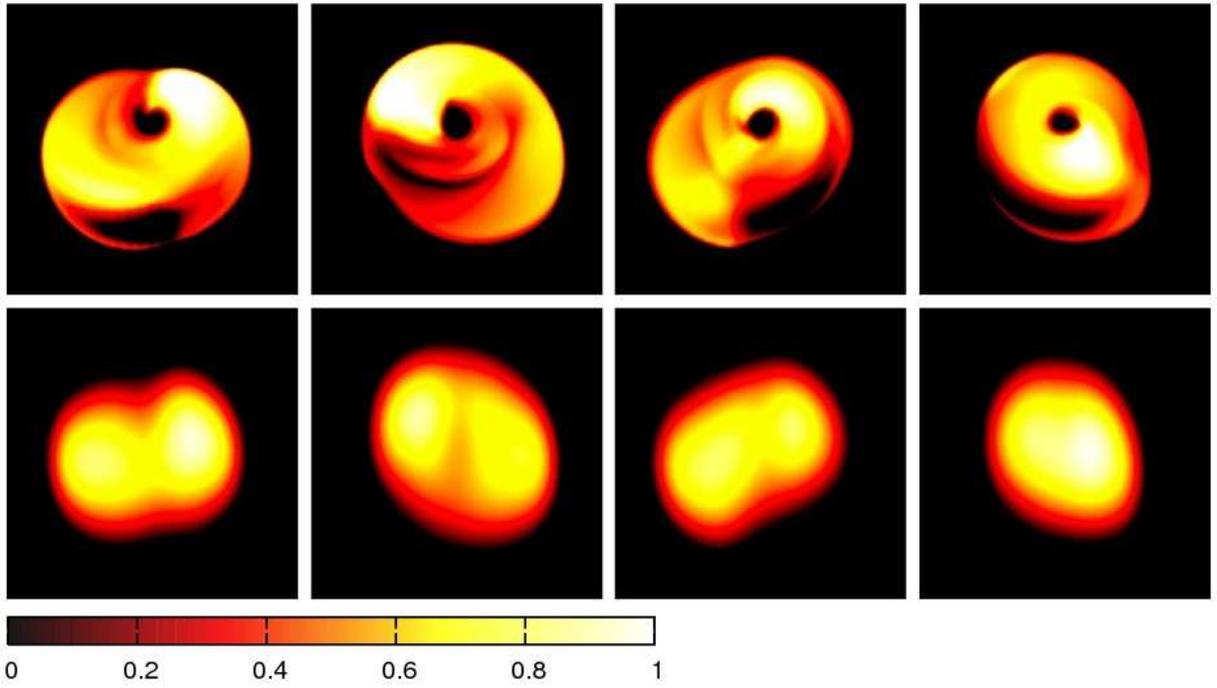}
\caption{
Same as the top two rows in Figure \ref{fig:image}, but with a Gaussian vertical density profile.
\label{fig:image2}}
\end{figure*}

\begin{figure*}
\epsscale{1.0}
\plotone{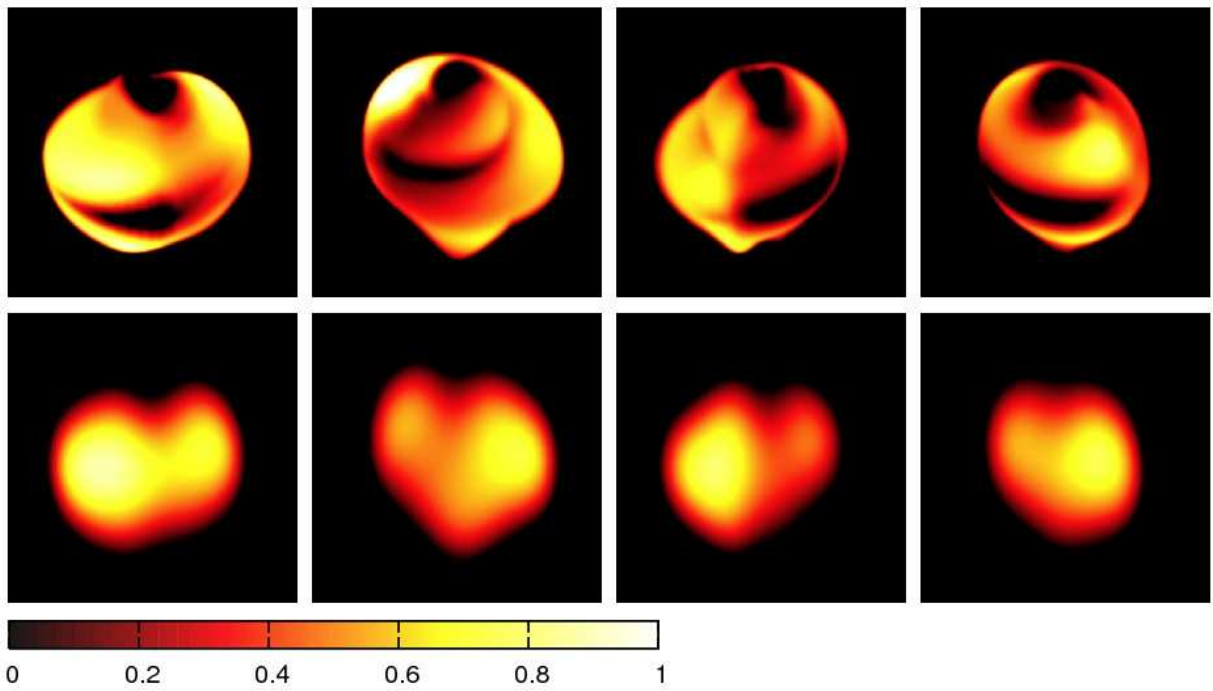}
\caption{
Same as the top two rows in Figure \ref{fig:image}, but with a viewing angle $i=45^\circ$.
\label{fig:image3}}
\end{figure*}

Figure~\ref{fig:image} shows the results of the observed intensity distributions ({\it top row}) and the images smeared by the spatial resolution of the {\it VSOP-2/ASTRO-G} ({\it middle row}) at 43 GHz. From left to right panels, we show the time evolution of the images at intervals of $\sim$ one month for the black hole in M87. The scale of the color bar is normalized in order that the time averaged images reproduce the observed spectrum as described in the previous section. The size of the images is $(80 GM/c^2)^2$, which corresponds $(3.8 \times 10^{16} cm)^2$ for M87. The spiral waves due to the SASI around the black hole in M87 evolve with spatial scales larger than the spatial resolutions of the {\it VSOP-2/ASTRO-G} observations. An exponential vertical density profile is assumed for the case shown in Figure~\ref{fig:image}. Thus, if shock waves exist in M87, as suggested by \cite{becker08} and \cite{mandal2008} based on the spectral fitting analysis, the time variation features will be detected by the {\it VSOP-2/ASTRO-G} satellite at 43 GHz. Almost all emission comes from the inside of the shocks, as a result of the density and the temperature peaks of the spiral waves. We also show the images for the Gaussian profile of the vertical density structure (Figure \ref{fig:image2}), and different viewing angle $i=45^\circ$ (Figure~\ref{fig:image3}) cases. As can be clearly seen in the bottom panels of these figures, we confirm the observational feasibility of SASI by the {\it VSOP-2/ASTRO-G}.


The most luminous parts of the intensity are predicted to exist at a distance of e.g. 30 $GM/c^2$ from the vicinity of black hole for some snap shots. This is in contrast to the calculations using the stationary accretion flow model where the most luminous part exists around the blue-shifted part of the accretion flow around the black hole shadow. This means that the position of the black hole in M87 can not be determined from the brightness distribution observed in a few months. In our simulations, such time-variable signatures will be smeared out by taking the average of the images with the time scale longer than the SASI timescale, i.e. several years in the case of M87. Before the launch of the {\it VSOP-2/ASTRO-G} satellite in 2013, the central regions of M87 will continue to be observed with existing telescopes at frequencies of 230 and 345 GHz, with spatial resolutions of 17 and 11 $\mu$as, respectively \citep{bl09}. It is considered that at 345 GHz, jet emission dominates over disk emission, as shown by \cite{bl09} (e.g. Figure~6). In Figure~\ref{fig:image}, we show that the effective emitting region at 345 GHz is smaller than that emitting at 43 GHz. This is because the accreting plasma is optically thin at 345 GHz and then the effective emitting region exists around the shadow. Only when the contamination of the emissions from the jets will not conceal the emission from the spiral waves due to SASI, it has a possibility that the time evolution features due to the SASI in the intensity map or the visibility will be detected by the interferometers at 345 GHz. However, our simulation shows that the SASI phenomena will be more clearly observed at 43 GHz because of the size of the effective emitting region is larger than the case of 345 GHz.

\section{Discussion and Summary}
 In this paper,
 we have investigated the time variable images and time-averaged spectra
 at 43 GHz frequency around a Schwarzschild black hole,
 employing general relativistic hydrodynamic and radiation transfer simulations.
 The Black Hole SASI causes the time variability
 and its effect most obvious near the black hole shadow.
 We applied our model to M87
 and all of our results are consistent with past observational data.
 The variability timescale around the core region,
 reported by \citet{harris1997} and also \citet{perlman2003},
 is naturally explained by our Black Hole SASI model.
 It should be noted that it does not mean to exclude
 the other steady disk models such as RIAF, since the radial profile
 of the time-averaged electron temperature for our model
 is similar to RIAF.

 As shown in this paper,
 the dynamical features in the vicinity of M87 black hole
 are expected to be captured by future space telescopes,
 such as the {\it VSOP-2/ASTRO-G} satellite.
 In steady state, low luminous accretion flow models,
 a broadband energy spectrum is expected \citep{yqn03,jk07,bl09,li2009}. 
 For more detailed comparisons with these broadband observational data,
 there are some improvements remaining to be made. 
 Firstly, the frame dragging effects directly affect the black hole shadow
 and time variability. 
 According to \cite{wang08} and \cite{li2009}, 
 TeV variability from M87 indicates a rapidly spinning black hole (see also \cite{jk07}).
 The dependence of Kerr parameters for the imaging
 are currently undertaken \citep{nagakura2010}.
 Secondly, our current hydrodynamic simulations
 do not treat the meridional plane consistently.
 As discussed in \citet{nagakura2009}, the steady solutions for
 axisymmetric shocked accretion flow in the meridional plane
 should be constructed. 
 Finally, the magnetic fields may play an important role
 for SASI dynamics.
 These effects have begun to attract attentions in recent years
 \citep{endeve2008}. Since it is well known that
 there are the steady shocked accretion solutions
 with magnetic field in black hole spacetime \citep{tmasaaki2006},
 we regard that the SASI could naturally take in place in these flows.
 In this respect, the effects of MRI to SASI should be investigated.
 The general relativistic magneto-hydrodynamic is strongly required
 in order to shed light on these issues.
 These efforts are also currently under way. 
 Last but not least, the Black Hole SASI model could provide an interpretation
 for time variability observed in other accreting black hole systems,
 such as galactic X-ray binaries and Sgr A*.


\acknowledgments 
 We would like to thank Shoichi Yamada for useful discussions and suggestions.
 We also thank Dr. Gandhi for valuable discussion and proofreading.
 This research is supported by the Grant-in-Aid for Scientific Research Fund of the Ministry of Education, Culture, Sports, Science and Technology, Japan [Young Scientists (B) 21740149 (RT)]. 

\end{document}